\documentclass[10pt]{article}

\usepackage{amssymb}

\newcommand{\beq}{\begin{equation}}
\newcommand{\eeq}{\end{equation}}
\newcommand{\bea}{\begin{eqnarray}}
\newcommand{\eea}{\end{eqnarray}}

\begin{document}
\pagestyle{empty}

\hfill hep-th/0508103

\begin{center}

\vspace*{50mm}
{\LARGE Dyonic Anomalies}

\vspace*{20mm}
{\large M{\aa}ns Henningson and Erik P. G. Johansson}

\vspace*{10mm}
Department of Fundamental Physics\\
Chalmers University of Technology\\
S-412 96 G\"oteborg, Sweden

\end{center}

\vspace*{20mm} \noindent
{\bf Abstract:}\\
We consider the problem of coupling a dyonic $p$-brane in $d  = 2 p + 4$ space-time dimensions to a prescribed $(p + 2)$-form field strength. This is particularly subtle when $p$ is odd. For the case $p = 1$, we explicitly construct a coupling functional, which is a sum of two terms: one which is linear in the prescribed field strength, and one which describes the coupling of the brane to its self-field and takes the form of a Wess-Zumino term depending only on the embedding of the brane world-volume into space-time. We then show that this functional is well-defined only modulo a certain anomaly, related to the Euler class of the normal bundle of the brane world-volume.

\newpage \pagestyle{plain}
\section{The problem}
In a $d$-dimensional space-time $M$, a $(p + 2)$-form field strength $h$ can couple electrically to a $p$-brane, or magnetically to a $(d - p -4)$-brane \cite{Teitelboim}. These two basic couplings are interchanged by electric-magnetic duality, which is not manifest in a Lagrangian formulation, though. 

When no magnetic charges are present, the electric coupling is introduced as an interaction term in the action
\beq
S_{el}^0 = q \int_\Sigma X^* (b) . \label{Wilson}
\eeq
Here $q \in \mathbb Z$ denotes the electric charge of a $p$-brane, $\Sigma$ is its $(p + 1)$-dimensional world-volume, the map
\beq
X : \Sigma \rightarrow M
\eeq
describes the embedding into space-time, and $b \in \Omega^{p + 1} (M)$ is a gauge potential for the field strength, i.e. $h = d b$.  

The magnetic coupling is introduced as source term in the Bianchi identity of the field strength:
\beq
d h = 2 \pi g \delta_X . \label{Bianchi}
\eeq
Here $g \in \mathbb Z$ denotes the magnetic charge of a $(d - p - 4)$-brane, and $\delta_X \in \Omega^{p + 3} (M)$ is the Poincar\'e-dual of (the image of the embedding into space-time of) its world-volume.

Next, we consider the case with both an electrically charged $p$-brane and  a magnetically charged $(d - p - 4)$-brane present. Generically, their world-volumes  do not intersect in $d$ dimensions. Outside the world-volume of the magnetically charged brane, the field strength $h$ is thus closed and may locally be derived from a gauge potential $b$, although we will have to introduce different such potentials related by gauge transformations on different patches of space-time. The electric coupling can then be constructed more or less in the same way as before by a partition of unity argument. 

The case when $d = 2 p + 4$ allows for more subtleties. A $(p + 2)$-form field strength $h$ can then couple to a dyonic $p$-brane carrying both electric and magnetic charges $(q, g)$. So here we cannot use the above argument of non-intersecting world-volumes for electrically and magnetically charged branes to define the couplings. However, for $p$ even, we can apply an electric-magnetic duality transformation, after which a previously dyonic $p$-brane will be purely electrically charged, and the above definition of the electric coupling term can be used. But when $p$ is odd, the situation cannot be simplified by an electric-magnetic duality transformation \cite{Deser-Gomberoff-Henneaux-Teitelboim}. 

One way of understanding the difference between the cases when $p$ is even or odd is to consider a situation with two dyonic branes with electric and magnetic charges $(q, g)$ and  $(q^\prime, g^\prime)$ respectively. In the presence of the second brane, the 'wave-function' of the first brane is not really a function, but rather a section of a non-trivial line-bundle over the configuration space of the second brane, i.e. the space of all configurations for which the two branes do not intersect. The Chern class of this line-bundle is proportional to the integer
\beq
n = q g^\prime - (-1)^p g q^\prime .
\eeq
We see that for $p$ even, $n$ would vanish when the two branes are identical, so the topological class of the wave-function of one brane is not affected by the presence of the other brane in this case. This is consistent with the above remark that for $p$ even, any single brane is equivalent under a duality transformation to a purely electrically charged brane. But for $p$ odd, we see that this integer is non-zero, so the topological class of the wave-function of a brane then depends on the presence of other branes of the same type. Since by local considerations one cannot really tell whether two branes are really distinct or just parts of a single brane, we must conclude that the nature of the wave-function of a single brane is ill-defined because of the presence of that brane itself. 

Another indication that something subtle might happen for $p$ odd is that in this case, the brane has an even-dimensional world-volume, thus allowing for an anomaly \cite{Brax-Mourad}\cite{Henningson}\cite{Berman-Harvey}. Indeed, rewriting the electric coupling term as
\beq
S_{el}^0 = q \int_M b \wedge \delta_X ,
\eeq
we see that because of the modified Bianchi identity for the field strength $h$, there is an anomaly inflow from space-time $M$ to the theory on the brane world-volume $\Sigma$. The corresponding anomaly polynomial is given by $q g$ times the pull-back of the Poincar\'e dual $\delta_X$ to the world-volume. (More precisely, one has to consider the pull-back to a two-parameter family of embeddings of world-volumes into space-time \cite{Atiyah-Singer}\cite{Zumino}.) By the Thom isomorphism, this pullback equals the Euler class $\chi (N)$ of the normal bundle $N$ of the world-volume $\Sigma$ embedded into space-time $M$ by the map $X$. This classical anomaly means that the purely bosonic sector of the theory is inconsistent in itself, and has to be completed with other degrees of freedom, e.g. chiral world-volume fermions that are sections of bundles associated to $N$, chosen so that the total quantum anomaly cancels.

The purpose of this note is to investigate these issues in more detail. The physically relevant cases are $p = 1$ (self-dual strings in six-dimensional $(2, 0)$ theory) and $p = 3$ ($D3$-branes in ten-dimensional type IIB string theory). For concreteness and simplicity, we will only consider the first of these cases, but we expect that the generalization to arbitrary odd $p$ should be straightforward, although more tedious. So henceforth $\Sigma$ denotes a two-dimensional string world-sheet embedded in a six-dimensional space-time manifold $M$ by a map $X$, and $h$ is a three-form field strength obeying the Bianchi identity (\ref{Bianchi}). Our aim is now to construct an electric coupling term $S_{el}$ as a functional of these data. As we have indicated above, this is not quite possible: The functional that we will construct is only well-defined modulo an anomaly, which however depends only on the embedding data of the brane world-volume, and thus may be cancelled by other contributions in a complete theory.

\section{The solution...}
At the locus of a magnetically charged brane, the gauge potential $b$ is not a meaningful concept. We therefore begin by rewriting the electric coupling (\ref{Wilson}) directly in terms of the field strength $h$. This can be accomplished by introducing an open three-manifold $D$, the boundary of which is given by the string world-sheet $\Sigma$, i.e. $\partial D = \Sigma$, and extending the embedding map $X$ to a map
\beq
\tilde{X} : D \rightarrow M ,
\eeq
i.e. the restriction of $\tilde{X}$ to $\Sigma$ equals $X$. We now define
\beq
S_{el}^0 = q \int_D \tilde{X}^* (h) .
\eeq
By Stokes' theorem, this definition of $S_{el}^0$ agrees with (\ref{Wilson}) in the absence of magnetically charged branes. In particular, the choice of extension $\tilde{X}$ of $X$, subject to the above conditions, is immaterial in this case.

But in the presence of magnetically charged branes, when $h$ is no longer a closed three-form, $S_{el}^0$ does depend on the choice of extension $\tilde{X}$ of $X$. To remedy this, we wish to add a term, denoted as $S_{WZ}$ for reasons that will be apparent later, that depends solely on $\tilde{X}$ (and not on the field strength $h$). This term should be constructed so that the total coupling term 
\beq
S_{el} = S_{el}^0 + q g S_{WZ} 
\eeq
depends only on $h$ and $X$, but not on the choice of $\tilde{X}$.

To construct such a term, we use the following geometric considerations: The embedding of $\Sigma$ in $M$ defines the normal bundle $N$ over $\Sigma$, which is endowed with a canonical connection. We also have the corresponding unit sphere bundle $S$ over $\Sigma$. The map $\tilde{X}$ defines a canonical section of the latter bundle, given by the unit tangent vector field of the embedding of $D$, which is orthogonal to the tangent vectors of the embedding of $\Sigma$. At each point $p \in \Sigma$, the fiber of $S$ is isomorphic in a non-canonical way to the three-sphere
\beq
S^3 = \{ v = (v^1, v^2, v^3, v^4) \in \mathbb R^4 | (v^1)^2 + (v^2)^2 + (v^3)^2 + (v^4)^2 = 1 \} .
\eeq
We now choose a specific isomorphism, i.e. a trivialization for the normal bundle $N$. The section of $S$ defined by the map $\tilde{X}$ then induces a map
\beq
v : \Sigma \rightarrow S^3 .
\eeq
This map can be extended to a map
\beq
\tilde{v} : D \rightarrow S^3 ,
\eeq
i.e. the restriction of $\tilde{v}$ to $\Sigma$ equals $v$.

Given a choice of trivialization of $N$ and a choice of extension $\tilde{v}$ of $v$ as described above, we may now define the gauged Wess-Zumino term
\beq
S_{WZ} = \frac{1}{6 \pi} \int_D \epsilon_{i j k l}  \tilde{v}^i \, d \tilde{v}^j \, d \tilde{v}^k \, d \tilde{v}^l + \frac{1}{8 \pi} \int_\Sigma \epsilon_{i j k l} \left( - 2 v^i \, d v^j \, A^{k l} + v^i \, A^{j k} \, A^{l m} v^m \right) .
\eeq
Here $A^{i j} = - A^{j i}$ are the one-forms representing the connection on $N$ with respect to the chosen trivialization. Note that the integrand of the first term is given by $2 \pi$ times the pullback by $\tilde{v}$ of the unit volume-form on $S^3$. By standard arguments, it can be shown that although the Wess-Zumino term is defined in terms of the extension $\tilde{v}$, it is really a well-defined functional modulo $2 \pi$ of the map $v$. Furthermore, as described above, the latter map is induced by the map $\tilde{X}$, once a trivialization of the normal bundle $N$ of $\Sigma$ is chosen. 

\section{...and its properties}
To write down the Wess-Zumino term in the previous section, we had to choose a trivialization of the normal bundle. We will now investigate the transformation properties of $S_{WZ}$ under an infinitesimal change of trivialization with parameters $\theta^{i j} = - \theta^{j i}$, acting as
\bea
\delta v^i & = & \theta^{i j} v^j \cr
\delta A^{i j} & = & \theta^{i m} A^{m j} - A^{i m} \theta^{m j} - d \theta^{i j} .
\eea
(The extension $\tilde{X}$ of the map $X$ is held fixed, though.) A short calculation shows that the Wess-Zumino term transforms as
\beq
\delta S_{WZ} = - \frac{1}{16 \pi} \int_\Sigma \epsilon_{i j k l} \theta^{i j} \, d A^{k l} , \label{variation}
\eeq
while $S_{el}^0$ of course is invariant. The remarkable property of $S_{WZ}$ is, that although it is not invariant under this transformation, its variation depends only on the connection one-forms $A$ and the parameters $\theta$ (and not on the map $v$) \cite{Witten}. 

We remark that $S_{WZ}$ is in fact equivalent to a gauged version of the $SU (2) \simeq S^3$ Wess-Zumino term that figures prominently in two-dimensional conformal field theory. This term has a global $SO (4) \simeq SU (2) \times SU (2)$ symmetry, which in our case corresponds to the structure group of the normal bundle $N$. As we have seen in the previous paragraph, it is not possible to gauge all of this symmetry, but only e.g. the diagonal $SU (2)$ subgroup.

Returning now to the variation (\ref{variation}), it is of a standard form that follows by applying the descent procedure to the Euler class density 
\beq
\chi = \epsilon_{i j k l} F^{i j} F^{k l} ,
\eeq
where  $F^{i j} = d A^{i j} + A^{i m} A^{m j}$ is the curvature two-form of the connection on $N$. Indeed, the four-form $\chi$ is closed and invariant under a change of trivialization. Locally (but in general not globally), it can be written as $\chi = d \Omega$ for 
\beq
\Omega = \epsilon_{i j k l} \left( A^{i j} d A^{k l} + \frac{2}{3} A^{i j} A^{k m} A^{m l} \right) .
\eeq
The variation of the three-form $\Omega$ under a change of trivialization is exact, $\delta \Omega = d {\cal A}$, for 
\beq
{\cal A} = - \epsilon_{i j k l} \theta^{i j} d A^{k l} .
\eeq
As we have seen above, the classical anomalous variation of the Wess-Zumino term is a multiple of ${\cal A}$. Consistency of the model requires that this classical variation be cancelled by  e.g. the quantum anomaly of world-volume chiral fermions that are sections of bundles associated to the normal bundle $N$. This mechanism works both for the self-dual string in six-dimensional $(2, 0)$ theory and the $D3$-brane in ten-dimensional type IIB theory. 

Apart from the anomaly found above, the complete electric coupling $S_{el} = S_{el}^0 + q g S_{WZ}$ can be regarded as a functional of the field strength $h$ and the extension $\tilde{X}$ of the embedding map $X$. It remains to show that $S_{el}$ is in fact independent of the choice of $\tilde{X}$, and thus induces a functional of $h$ and $X$ (again neglecting its anomalous dependence on the choice of trivialization of the normal bundle $N$). To see this, we note that the modified Bianchi identity implies that the integral of $h$ over a three-sphere that links the string world-sheet equals $2 \pi  g$. Similarly, the integral of $\frac{1}{6 \pi} \epsilon_{i j k l}  \tilde{v}^i \, d \tilde{v}^j \, d \tilde{v}^k \, d \tilde{v}^l$ over a three-sphere around the origin equals unity. So both of these terms have a kind of singularity at the origin. But these singularities cancel in the combined integrand of $S_{el}$
\beq
q \tilde{X}^* (h) - q g \frac{1}{6 \pi} \epsilon_{i j k l}  \tilde{v}^i \, d \tilde{v}^j \, d \tilde{v}^k \, d \tilde{v}^l  ,
\eeq
which therefore can be extended to a closed three-form over the origin. By Stokes' theorem, the integral of this form over $D$ can then be rewritten as an integral over the boundary $\Sigma$, and is thus indeed independent of the choice of extension $\tilde{X}$ over $D$.

For a more explicit demonstration of the independence of $\tilde{X}$, one may proceed as follows: It is straightforward  to compute the variation of $S_{WZ}$ under a change of extension $\tilde{X}$ (which induces a variation of the map $\tilde{v}^i$ but leaves the connection one-forms $A^{i j}$ invariant). To compute the variation of $S_{el}^0$, one has to regularize the Poincar\'e dual $\delta_X$ that appears in the Bianchi identity of the field strength $h$ by replacing it with another closed form, constructed as a bump-form in the radial direction times an angular three-form. (To write down the latter form, one needs to choose a trivialization of the normal bundle $N$, just as one does to define $S_{WZ}$.) One then finds that the variations of $S_{WZ}$ and $S_{el}^0$ cancel, so that $S_{el}$ is indeed invariant.

\vspace*{5mm}
M.H. is a Research Fellow at the Royal Swedish Academy of Sciences (KVA). He thanks the Benasque Center for Science for its nice atmosphere, where this work was partly done.

\end{document}